\documentclass{Interspeech2024}
\usepackage{subcaption}

\usepackage{hyperref}
\usepackage{url}
\usepackage{booktabs}       
\usepackage{amsfonts}       
\usepackage{nicefrac}       
\usepackage{microtype}      
\usepackage{xcolor}         
\global\definecolor{darkgray}{gray}{0.3}
\definecolor{ired}{RGB}{255,59,48}
\global\definecolor{iorange}{RGB}{255,149,0}
\definecolor{iyellow}{RGB}{255,204,10}
\definecolor{igreen}{RGB}{52,199,89}
\definecolor{iblue}{RGB}{0,122,255}
\definecolor{ipurple}{RGB}{175,82,222}
\usepackage{xspace}    
\usepackage{tabularx}         
\usepackage{amstext} 
\usepackage{array}   
\usepackage{algorithm,algorithmicx,algpseudocode}

\usepackage{pgfplots}
\usepackage{amsmath} 
\usepackage{amssymb} 

\usepackage{tikz}

\usepackage{pgfplots, pgfplotstable}
\pgfplotsset{compat=1.18}
\usetikzlibrary{calc, shapes}

\usepackage{amsmath,amsfonts,bm}

\def\eqref#1{equation~\ref{#1}}

\def\1{\bm{1}}

\def\vh{{\bm{h}}}

\def\vx{{\bm{x}}}

\DeclareMathAlphabet{\mathsfit}{\encodingdefault}{\sfdefault}{m}{sl}
\SetMathAlphabet{\mathsfit}{bold}{\encodingdefault}{\sfdefault}{bx}{n}

\newcolumntype{C}{>{$}c<{$}} 
\newcommand{\tableresultsablation}{
\begin{table*}[th]
  \caption{Evaluation results of the models in the ablation study, with bold numbers indicating the better value between the modified versions and the baseline method.}
  \label{tab:result_ablation}
  \centering
  \begin{tabular}{l |CCCCCCC}
    \toprule
    Model & \text{MAE }\downarrow & \text{M-STFT }\downarrow & \text{PESQ }\uparrow & \text{MCD }\downarrow & \text{V/UV F$1$ }\uparrow &     \text{Periodicity }\downarrow & \text{Pitch }\downarrow \\
    \midrule
    Discrete dist. (Baseline) & \bm{0.3337} & \bm{1.227} & \bm{2.493} & \bm{1.505}  & \bm{0.9407} & \bm{0.1527} & \bm{35.63} \\
    Uniform dist.  & 0.3415 & 1.252 & 2.444 & 1.544 & \bm{0.9407} & 0.1541 & 37.59 \\
    Normal dist.  & 0.3369 & 1.247 & 2.412 & 1.542 & 0.9382 & 0.1617 & 37.61 \\
        \specialrule{0.01em}{0.01em}{0.01em}
    \midrule
    Baseline & 0.3337 & \bm{1.227} & \bm{2.493} & \bm{1.505}  & \bm{0.9407} & \bm{0.1527} & \bm{35.63} \\
    + async shift & \bm{0.3297} & 1.237 & 2.437 & 1.554 & 0.9403 & 0.1535 & 37.25  \\
        \specialrule{0.01em}{0.01em}{0.01em} 
    \midrule
   Baseline & 0.3337 & \bm{1.227} & \bm{2.493} & 1.505  & \bm{0.9407} & \bm{0.1527} & \bm{35.63} \\
   JenGAN for $G$ only & 0.3434 & 1.245 & 2.369 & 1.523 & 0.9391 & 0.1576 & 43.18  \\
   JenGAN for $D$ only & \bm{0.3306} & 1.238 & 2.380 & \bm{1.494} & 0.9383 & 0.1601 & 42.26  \\
        \specialrule{0.01em}{0.01em}{0.01em} 
    \bottomrule
  \end{tabular}
\end{table*}
}

\newcommand{\tableresultsbaseline}{
\begin{table*}[th]
  \caption{Evaluation results of the original HiFi-GAN model, the application of PhaseAug \cite{phaseaug}, and the application of JenGAN. The better value is highlighted in bold.}
  \label{tab:result_baseline}
  \centering
  \begin{tabular}{l |CCCCCCC}
    \toprule
    Model & \text{MAE }\downarrow & \text{M-STFT }\downarrow & \text{PESQ }\uparrow & \text{MCD }\downarrow & \text{V/UV F$1$ }\uparrow &     \text{Periodicity }\downarrow & \text{Pitch }\downarrow \\
    \midrule
    HiFi-GAN (100\%) & 0.2237 & 1.006 & 3.628 & 1.151 & 0.9624 & 0.1063 & 25.17  \\
    + PhaseAug  & 0.2122 & 1.002 & 3.666 & 1.137 & 0.9650 & \bm{0.0995} & 24.81  \\
    + JenGAN  & \bm{0.2107} & \bm{1.000} & \bm{3.687} & \bm{1.124} & \bm{0.9677} & 0.1027 & \bm{23.85}  \\
        \specialrule{0.01em}{0.01em}{0.01em}
    \midrule
    HiFi-GAN (10\%) & 0.2413 & 1.037 & 3.518 & 1.252 & 0.9602 & 0.1085 & 25.70 \\
    + PhaseAug  & \bm{0.2216} & \bm{1.012} & \bm{3.606} & \bm{1.223} & \bm{0.9613} & \bm{0.1074} & 26.81 \\
    + JenGAN  & 0.2352 & 1.027 & 3.577 & 1.241 & 0.9611 & 0.1080 & \bm{25.48} \\
        \specialrule{0.01em}{0.01em}{0.01em} 
    \midrule
   HiFi-GAN (1\%) & 0.3740 & 1.276 & 2.162 & 1.610 & 0.9314 & 0.1705 & 42.95 \\
    + PhaseAug  & \bm{0.3377} & \bm{1.189} & \bm{2.681} & \bm{1.471} & \bm{0.9416} & \bm{0.1503} & 37.56 \\
    + JenGAN & \bm{0.3337} & 1.227 & 2.493 & 1.505 & 0.9407 & 0.1527 & \bm{35.63} \\
        \specialrule{0.01em}{0.01em}{0.01em} 
    \bottomrule
  \end{tabular}
\end{table*}
}

\newcommand{\tableresultsother}{
\begin{table*}[th]
  \caption{Evaluation results of different vocoder models, including Avocodo \cite{avocodo} and BigVGAN-base \cite{bigvgan}, with and without the JenGAN algorithm. The better value is highlighted in bold, indicating the improvements achieved by using the JenGAN algorithm.}
  \label{tab:result_other}
  \centering
  \begin{tabular}{l |CCCCCCC}
    \toprule
    Model & \text{MAE }\downarrow & \text{M-STFT }\downarrow & \text{PESQ }\uparrow & \text{MCD }\downarrow & \text{V/UV F$1$ }\uparrow &     \text{Periodicity }\downarrow & \text{Pitch }\downarrow \\
    \midrule
    Avocodo & 0.1907 & 0.9657 & 3.767 & 1.057 & 0.9652 & 0.1009 & 23.76  \\
    + JenGAN  & \bm{0.1789} & \bm{0.9552} & \bm{3.817} & \bm{1.042} & \bm{0.9678} & \bm{0.09687} & \bm{21.94}  \\
        \specialrule{0.01em}{0.01em}{0.01em} 
    \midrule
    BigVGAN-base & 0.1705 & 0.9477 & 3.758 & 1.005 & \bm{0.9646} & 0.1033 & 23.05  \\
    + JenGAN  & \bm{0.1606} & \bm{0.9458} & \bm{3.789} & \bm{0.9945} & \bm{0.9646} & \bm{0.1011} & \bm{22.63}  \\
        \specialrule{0.01em}{0.01em}{0.01em}
    \bottomrule
  \end{tabular}
\end{table*}
}
\newcommand{\algorithmgen}{
\begin{algorithm}[!ht]
    \algrenewcommand\algorithmicindent{1.0em}
  \caption{JenGAN algorithm (HiFi-GAN Generator)} \label{alg:JenGAN_method_gen}
  \hspace*{\algorithmicindent} \textbf{Original block} $M(x)$ \\
  \hspace*{\algorithmicindent} \textbf{Input}: intermediate signal $\vx[n]$
  \begin{algorithmic}[1]
      \State Compute $\vh[n] \leftarrow  M(\vx[n])$
  \end{algorithmic}
  \hspace*{\algorithmicindent} \\
  \hspace*{\algorithmicindent} \textbf{JenGAN block} \\
  \hspace*{\algorithmicindent} \textbf{Input}: signal $\vx[n]$, upsample rate $r$, shifting value $\delta$
  \begin{algorithmic}[1]
      \State Compute $\vx_{\delta/r}[n] \leftarrow \vx[n] \ast F_{\sinc}(- \delta / r)$
      \State Compute $\vh_{\delta}[n] \leftarrow M(\vx_{\delta/r}[n])$  
      \State Compute $\vh[n] \leftarrow \vh_{\delta}[n] \ast F_{\sinc}( + \delta)$ 
  \end{algorithmic}
\end{algorithm}
}

\newcommand{\algorithmdis}{
\begin{algorithm}[!ht]
    \algrenewcommand\algorithmicindent{1.0em}
  \caption{JenGAN algorithm (HiFi-GAN Discriminator)} \label{alg:JenGAN_method_dis}
  \hspace*{\algorithmicindent} \textbf{Original block} $M(x)$ \\
  \hspace*{\algorithmicindent} \textbf{Input}: intermediate signal $\vx[n]$
  \begin{algorithmic}[1]
      \State Compute $\vh[n] \leftarrow M(\vx[n])$
      \State Append $\vh[n]$ to $\mathrm{fmap}$  
  \end{algorithmic}
  \hspace*{\algorithmicindent} \\
  \hspace*{\algorithmicindent} \textbf{JenGAN block} \\
  \hspace*{\algorithmicindent} \textbf{Input}: signal $\vx[n]$, downsample rate $r$, shifting value $\delta$
  \begin{algorithmic}[1]
      \State Compute $\vx_{\delta}[n] \leftarrow \vx[n] \ast F_{\sinc}( - \delta)$
      \State Compute $\vh_{\delta/r}[n] \leftarrow M(\vx_{\delta}[n])$
      \State Compute $\vh[n] \leftarrow \vh_{\delta/r}[n] \ast F_{\sinc}( + \delta / r)$
      \State Append $\vh[n]$ to $\mathrm{fmap}$  
  \end{algorithmic}
\end{algorithm}
}
\DeclareMathOperator{\sinc}{sinc}

\usepackage{enumitem}
\let\Phi\varPhi

\makeatletter
\DeclareRobustCommand\onedot{\futurelet\@let@token\@onedot}
\def\@onedot{\ifx\@let@token.\else.\null\fi\xspace}

\def\etal{\emph{et al}\onedot}
\makeatother

\interspeechcameraready

\title{JenGAN: Stacked Shifted Filters in GAN-Based Speech Synthesis}

\name[affiliation={1 \ast \dagger}]{Hyunjae}{Cho}
\name[affiliation={2 \ast \dagger}]{Junhyeok}{Lee}
\name[affiliation={3 \dagger}]{Wonbin}{Jung}

\address{
  $^1$Seoul National University (SNU), Republic of Korea\\
  $^2$Supertone Inc., Republic of Korea\\
  $^3$Korea Advanced Institute of Science and Technology (KAIST), Republic of Korea}
\email{choing84@snu.ac.kr, jun.hyeok@supertone.ai, santabin0222@gamil.com}


\begin{document}

\maketitle
\begin{abstract}
Non-autoregressive GAN-based neural vocoders are widely used due to their fast inference speed and high perceptual quality.
However, they often suffer from audible artifacts such as tonal artifacts in their generated results.
Therefore, we propose JenGAN, a new training strategy that involves stacking shifted low-pass filters to ensure the shift-equivariant property.
This method helps prevent aliasing and reduce artifacts while preserving the model structure used during inference.
In our experimental evaluation, JenGAN consistently enhances the performance of vocoder models, yielding significantly superior scores across the majority of evaluation metrics.
\end{abstract}

\def\thefootnote{*}\footnotetext{Equal contribution}
\def\thefootnote{\textdagger}\footnotetext{Work done at maum.ai Inc.}
\def\thefootnote{\arabic{footnote}}

\section{Introduction}

Neural vocoders take a sequence of audio features, such as mel-spectrograms, and generate an audio waveform as an output.
Autoregressive vocoders \cite{wavenet, wavernn}, which generate waveform samples by conditioning on previous samples, are known to generate relatively high-quality audio signals.
Despite their high-quality outputs, the slow performance of autoregressive vocoders has prompted research into non-autoregressive alternatives in neural vocoder models.
Due to the complexity of neural vocoders in transforming spectrograms into waveforms, it is not sufficient to rely solely on the mel-spectrogram reconstruction loss function, especially for the non-autoregressive model.
Therefore, non-autoregressive vocoders incorporate techniques based on generative adversarial networks (GANs) \cite{gan, hifi-gan, univnet, avocodo}, normalizing flows \cite{Parallel_wavenet, waveglow}, and diffusion models \cite{wavegrad, diffwave} to overcome the complexity.
Among these techniques, GAN-based vocoders have become popular for their fast generation speed and ability to produce audio of acceptable quality.

Previous researches \cite{avocodo, upsample_alias} suggest that the upsampling and downsampling layers in GAN-based vocoders can contribute to the generation of audible artifacts.
Pons \etal{} \cite{upsample_alias} have observed that using transposed convolution layers for upsampling in HiFi-GAN \cite{hifi-gan} can result in the occurrence of audible artifacts, such as tonal artifacts.
In addition, signal aliasing was observed in downsampled waveforms in the discriminators of GAN-based vocoders.
According to the Nyquist–Shannon sampling theorem \cite{Nyquist}, downsampling layers are prone to cause aliasing if the signal has not been band-limited before downsampling.
Furthermore, pointwise nonlinearity activation functions have been identified as sources of aliasing in discrete signals \cite{sytlegan3, bigvgan}.
Karras \etal{} \cite{sytlegan3} propose a solution that involves upsampling the signal prior to the nonlinearity activation function and subsequently downsampling it to mitigate aliasing.
However, this approach leads to increased memory requirements and slower generation and training speeds due to the inclusion of additional upsampling and downsampling layers.

Recently, several studies in speech synthesis have suggested applying differentiable digital signal processing (DDSP) \cite{ddsp, NSF, pwghnp, nansypp} or discrete audio codecs \cite{soundstream,encodec}.
DDSP-applied studies adopt the source-filter theory \cite{sourcefilter} to model human speech as being composed of a harmonic-related source and excitation-related filter with differentiable features.
On the other hand, neural audio codecs encode speech into discrete tokens by residual vector quantization and resynthesize speech from quantized vectors. 
Given that these studies are closely related to the vocoding task, we have chosen not to compare these approaches. 
Instead, our goal is to explore the potential of HiFi-GAN, one of the most renowned architectures for various applications, without making any modifications to its architecture.

This paper introduces JenGAN, an anti-aliasing algorithm designed to reduce audible artifacts in GAN-based vocoders.
JenGAN incorporates two convolution operations with sinc kernel before and after the original neural network blocks to achieve shift-equivariant property, which is related to aliasing \cite{sytlegan3}.
While prior studies \cite{sytlegan3, bigvgan} have recommended architecture modification to achieve anti-aliasing and shift-equivariance, JenGAN distinguishes itself by focusing solely on modifications to the training strategy.
Furthermore, the proposed method preserves the model's architecture during inference, leading to fast inference speeds and facilitating convenient fine-tuning.
Experimental results demonstrate that applying JenGAN enhances evaluation metric scores and yields more natural-sounding speech.

\section{Method}
\subsection{Stacked Shifted Filters}
When considering a specific block of the vocoder, discrete input and output signals of the block can be viewed as samples obtained from continuous signals.
Considering an ideal function operating on continuous signals, it is free from aliasing and can be perceived as a generalization of the original discrete block.
However, aliasing can occur when converting from the continuous domain to the discrete domain if the frequencies of the continuous signals exceed half of the sampling rate \cite{Nyquist}.
To prevent aliasing, it is necessary to limit the frequency of signals in the continuous representation to half of the sampling rate.
By adding a low-pass filter to both the input and output signals of a specific block, we can limit the frequency range of the signals within that block \cite{sytlegan3}.

\begin{figure}[th!]
  \centering
  \includegraphics[width=\linewidth]{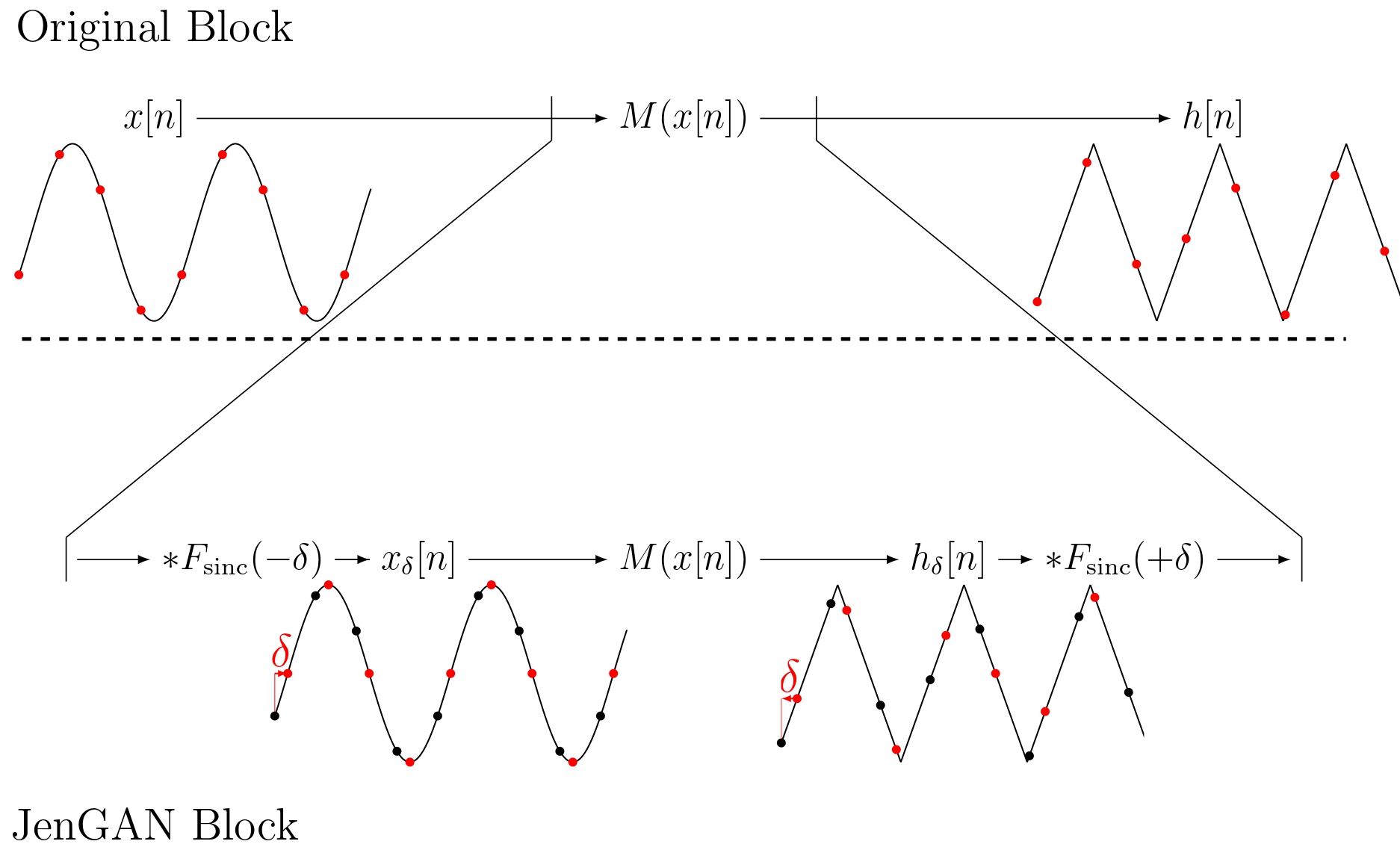}
\caption{This figure shows the overview of the JenGAN method. We shift the signal in the input of the block by $\delta$ and in the output of the block by $-\delta$ to achieve the shift-equivariant property.}
\label{fig:fig}
\end{figure}

However, using only a naive low-pass filter in the discrete representation to limit the frequency of signals does not accurately translate to frequency control in the continuous domain.
To address this issue, we leverage the shift-equivariant property, which is satisfied in ideal blocks in the continuous domain and their discretized model.
The shift-equivariant property ensures that when input signals are shifted by a specific real-valued number $\delta$ and output signals are subsequently shifted by $-\delta$, the resulting block remains the same in the continuous domain.
By training the block on the discrete representation using the shifting method and varying the values of $\delta$, we can approximate its behavior to that of the ideal block in the continuous domain, constraining its high-frequency bands.
This training method enables the blocks in the discrete domain to achieve shift-equivariant property, thereby resulting in a reduction of aliasing.

\subsection{JenGAN}
We propose JenGAN, an anti-aliasing strategy, in which the structure of the block is modified during the training process but remains unchanged during the inference.
We modify the block, $M(\vx[n])$, in the vocoders to:
\begin{equation}
    M\left(\vx[n] \ast F_{\sinc}( - \delta)\right) \ast F_{\sinc}( + \delta),
\end{equation}
where $\ast$ is a channel-wise convolution operator in the discrete domain, and $F_{\sinc}(\delta)$ is a shifted sinc filter defined over the range $-12\leq n \leq 12$, where $n \in \mathbb{Z}$ as follows:
\begin{equation}
    F_{\sinc}(\delta)[n] = \begin{cases}
    1 & \text{if } n+\delta =0 \\
    \sin(\pi(n+\delta))/(\pi(n+\delta)) & \text{otherwise } 
    \end{cases}
\end{equation}
This operation is implemented using PyTorch's $\mathrm{F.Conv1d}$\footnote{\url{https://pytorch.org/docs/stable/generated/torch.nn.functional.conv1d.html}}, employing a calculated sinc filter and setting the `groups' parameter equal to the input channel size.
During inference, we set $\delta$ to zero, guaranteeing that the block remains unchanged from its original form.
This modified block is identical to shifting the input and output representation in the time domain, and stacking this block and giving reconstruction loss is identical to forcing the model to learn shift equivariance.

If the block contains upsampling or downsampling layers, the sampling frequency of output signals will be different from the input sampling frequency by the upsampling or downsampling rate.
However, the magnitudes of the shifting value should remain consistent for both input and output signals in the continuous domain.
Therefore, we shift the output sinc filter by $r_d \cdot \delta$, where $r_d$ represents a frequency difference ratio.
So, we modify the block, denoted as $M(\vx[n])$, which includes upsampling or downsampling layers in the vocoders to:
\begin{equation}
    M(\vx[n] \ast F_{\sinc}(- \delta)) \ast F_{\sinc}(+ r_d \cdot \delta).
\end{equation}



Since the convolution function is differentiable, the proposed strategy enables gradient flow.
To select $\delta$, we randomly sample a value from a specific distribution, as described in Section \ref{Sampling method}.
Additionally, we consider the maximum value of the sampling value due to the finite length of the sinc filter used.
We utilize the sampling value for output shifting when upsampling blocks are included and for input shifting when downsampling blocks are included, ensuring the shifting value in the sinc filter is bounded. 
This process is applied to all blocks within the model.
The exact procedure for the JenGAN strategy in the generator is shown in Algorithm \ref{alg:JenGAN_method_gen}.

\algorithmgen{}
\algorithmdis{}

In GAN-based vocoder discriminators, the feature-matching loss is computed by aggregating the output signals from each block and comparing the aggregated values of the real and generated signals.
For accurate calculation of the feature-matching loss, we ensure that the identical shifting value is used to extract the intermediate feature in each block for every pair of real and generated signals.
Algorithm \ref{alg:JenGAN_method_dis} describes the detailed procedure for the JenGAN strategy in the discriminators.



\subsection{Shifting Value Sampling Method}
\label{Sampling method}
In both Algorithm \ref{alg:JenGAN_method_gen} and Algorithm \ref{alg:JenGAN_method_dis}, we employ three different sampling methods for the shifting value $\delta$: uniform distribution sampling from the range $[-2, 2)$, normal distribution sampling with a standard deviation of 2, and equal probability discrete sampling from the set of values $\{-2, -1, 0, 1, 2\}$.
For the baseline method, we use the discrete distribution sampling method for both generator and discriminators.

\tableresultsbaseline{}
\tableresultsablation{}
\section{Experiments}
\subsection{Dataset}

We trained all models on LJSpeech \cite{ljspeech}, which is a dataset containing $24$ hours of single-speaker speeches.
We divided the dataset into training and validation sets.
We trained vocoder models using $100\%$, $10\%$, and $1\%$ of the dataset to evaluate the effect of JenGAN on different dataset sizes.

\subsection{Baseline Method}
To evaluate JenGAN, we used HiFi-GAN\footnote{\url{https://github.com/jik876/hifi-gan}} \cite{hifi-gan} as our testing framework.
We regard the pair of MRF module \cite{hifi-gan} and ConvTranspose layer as a single block that applies the JenGAN method and implemented Algorithm \ref{alg:JenGAN_method_gen} and Algorithm \ref{alg:JenGAN_method_dis} to reduce the aliasing.
During training, we maintained all configurations consistent with the official HiFi-GAN $V1$ setup, with the exception of the JenGAN training strategy and dataset size.
Additionally, we applied the PhaseAug \cite{phaseaug} during training and compared its performance improvement against the JenGAN baseline method.
When training with $100\%$ and $10\%$ of the train split, we trained the model until $2.5$M steps.
For the $1\%$ models, we stopped training the models before their validation mel-spectrogram errors converged.
The convergence in the $1\%$ models typically occurred before within the first $100$k steps.
For the $1\%$ model, we selected a model with the minimum mel-spectrogram mean average error (MAE) among all checkpoints.
We trained all models from scratch on a V$100$ GPU.

\subsection{Ablation Study}
We conducted an ablation study on JenGAN by comparing its performance with models that had a single modification to the baseline method, to analyze the contribution of each component of JenGAN.
First, we evaluated the shifting value sampling strategy by comparing three different methods: sampling from a discrete distribution (baseline method), a uniform distribution, and a normal distribution.
Next, we asynchronously sampled different shifting values in the discriminators for each pair of real and generated signals.
At last, we compared the models that exclusively applied the JenGAN algorithm in each generator or discriminator.
For the ablation study, we only trained HiFi-GAN with a dataset size of $1\%$.

\tableresultsother{}

\subsection{JenGAN on Other Vocoder Models}
We also applied the JenGAN method to other vocoder models, including Avocodo \cite{avocodo} and BigVGAN-base \cite{bigvgan}.
The original Avocodo vocoder was trained using an unofficial open-source implementation\footnote{\url{https://github.com/rishikksh20/Avocodo-pytorch}} for a total of $3$M steps. 
When applying the JenGAN method, we begin training from the pre-trained Avocodo vocoder model at $2$M steps and continue for an additional $1$M steps.
For the BigVGAN-base vocoder, we utilized the official open-source implementation\footnote{\url{https://github.com/NVIDIA/BigVGAN}}.
We trained both the original model and the model with the JenGAN applied for a total of $1$M steps from scratch.
We trained all models using a dataset size of $100\%$ on a V$100$ GPU.

\subsection{Evaluation Metrics}
To evaluate our methods, we measured MAE,
multi-resolution STFT (M-STFT)\footnote{\url{https://github.com/csteinmetz1/auraloss}} \cite{parallelwavegan},
$16$kHz wide-band perceptual evaluation of speech quality (PESQ)\footnote{\url{https://github.com/ludlows/python-pesq}} \cite{pesq}, 
mel-cepstral distortion (MCD)\footnote{\url{https://github.com/ttslr/python-MCD}} \cite{mcd}, F$1$ score of voiced/unvoiced classification (V/UV F$1$)\footnote{\url{https://github.com/descriptinc/cargan}} \cite{CARGAN}, periodicity error \cite{CARGAN}, and pitch error \cite{CARGAN} as objective metrics.

\section{Results and Discussion}
\subsection{Baseline Method}
Table \ref{tab:result_baseline} presents the evaluation results of the original HiFi-GAN model, along with the performance when applying PhaseAug and JenGAN algorithms.
The evaluation results demonstrate that both the models trained with JenGAN and PhaseAug outperform the default HiFi-GAN.
The models with PhaseAug perform better than the models with JenGAN when training the models on small datasets, such as $1\%$ or $10\%$.
However, when training the model on the full dataset ($100\%$), applying the JenGAN algorithm outperforms applying the PhaseAug method.

\subsection{Ablation Study}
Table \ref{tab:result_ablation} displays the evaluation results of the models in the ablation study.
Based on the results, the discrete distribution sampling method achieves the best performance across most of the evaluation metrics.
This outcome suggests that applying the $F_{\sinc}$ function to both the input and output may have contributed to instability in the model.
In discrete distribution sampling methods, typically only one of the input or output undergoes the $F_{\sinc}$ function application.
When applying the async shift method, which involves shifting different amounts asynchronously in each pair of real and generated signals, the performance decreases compared to the baseline method.
Furthermore, when using only Algorithm \ref{alg:JenGAN_method_gen} in the generator or using only Algorithm \ref{alg:JenGAN_method_dis} in the discriminators, lower performances are observed compared to the baseline method.

\subsection{JenGAN on Other Vocoder Models}
Table \ref{tab:result_other} displays the evaluation results of two vocoder models, Avocodo \cite{avocodo} and BigVGAN-base \cite{bigvgan}.
The table presents the evaluation outcomes for both the original vocoder and the vocoder with the JenGAN algorithm applied.
The results indicate that applying the JenGAN algorithm to both the Avocodo and BigVGAN-base vocoder models leads to substantial improvements in the quality of the generated speech.
These findings suggest that the JenGAN method positively influences the performance of vocoder models, resulting in improved speech synthesis quality.

\subsection{Improved Harmonics}
\begin{figure}[!ht]
\setlength{\textfloatsep}{0pt}%
    \setlength{\textfloatsep}{0pt}%
    \setlength{\intextsep}{0pt}
    \centering
\includegraphics[width=\linewidth]
{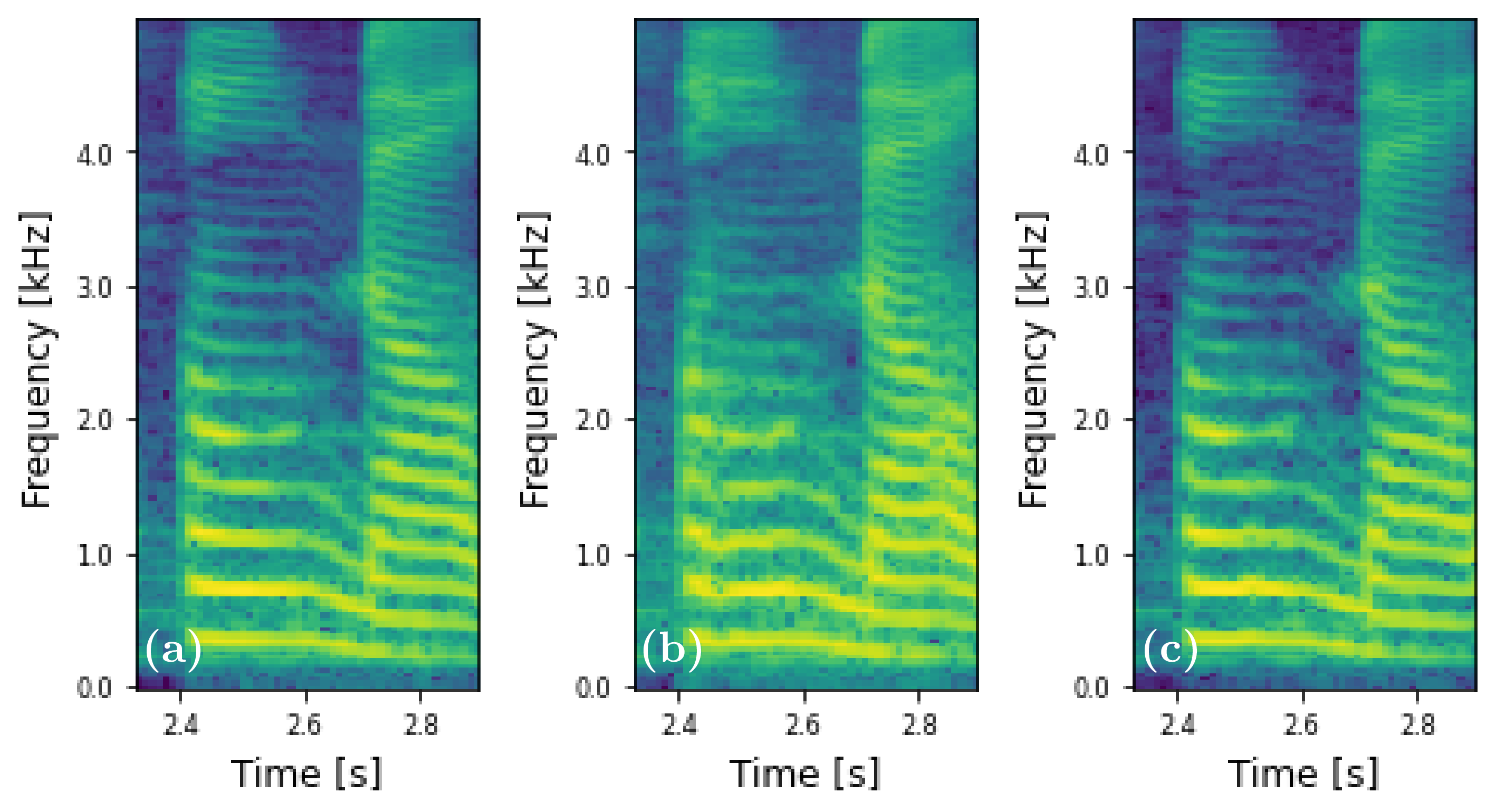}
\caption{Mel-spectrograms of (a) ground truth speech and speeches generated by (b) the original HiFi-GAN model, (c) the model applying JenGAN.}
\label{fig:baseline_image}
\end{figure}

Figure \ref{fig:baseline_image} shows the mel-spectrogram comparisons of (a) ground truth speech and speeches generated by (b) the original HiFi-GAN model and (c) the model with the JenGAN applied.
Both models are trained using the full dataset.
The results indicate that applying the JenGAN enhances the clarity of patterns in the mel-spectrogram images, resulting in the generation of more natural-sounding human harmonics.
This observation is also supported by high scores on the pitch evaluation metric when applying the JenGAN.

\section{Conclusions}
This paper describes JenGAN, a new strategy for training vocoder models that leverages the stacks of shifted low-pass filters.
This approach effectively mitigates the aliasing problem and reduces the artifacts that could manifest in the generated speeches.
Based on the results of the conducted experiments, the application of the JenGAN method to the vocoder models yields significant enhancement in the quality of the output speeches.
Furthermore, maintaining the model structure during inference steps provides the advantage of adaptability to a broad spectrum of vocoder models and convenient fine-tuning capabilities.

\bibliographystyle{IEEEtran}
\bibliography{jengan}

\end{document}